\documentclass[12pt]{article}
\usepackage{latexsym}
\usepackage{amsmath,amsfonts}
\usepackage{times}
\allowdisplaybreaks[4]
\catcode`\@=11

\newcount\hour
\newcount\minute
\newtoks\amorpm \hour=\time\divide\hour by 60\minute
=\time{\multiply\hour by 60 \global\advance\minute by-\hour}
\edef\standardtime{{\ifnum\hour<12 \global\amorpm={am}%
        \else\global\amorpm={pm}\advance\hour by-12 \fi
        \ifnum\hour=0 \hour=12 \fi
        \number\hour:\ifnum\minute<10
        0\fi\number\minute\the\amorpm}}
\edef\militarytime{\number\hour:\ifnum\minute<10 0\fi\number\minute}

\def\draftlabel#1{{\@bsphack\if@filesw {\let\thepage\relax
   \xdef\@gtempa{\write\@auxout{\string
      \newlabel{#1}{{\@currentlabel}{\thepage}}}}}\@gtempa
   \if@nobreak \ifvmode\nobreak\fi\fi\fi\@esphack}
        \gdef\@eqnlabel{#1}}
\def\@eqnlabel{}
\def\@vacuum{}
\def\marginnote#1{}
\def\draftmarginnote#1{\marginpar{\raggedright\scriptsize\tt#1}}
\def\draft{
        \pagestyle{plain}
        \overfullrule=2pt
        \oddsidemargin -.5truein
        \def\@oddhead{\sl \phantom{\today\quad\militarytime} \hfil
        \smash{\Large\sl DRAFT} \hfil \today\quad\militarytime}
        \let\@evenhead\@oddhead
        \let\label=\draftlabel
        \let\marginnote=\draftmarginnote
        \def\ps@empty{\let\@mkboth\@gobbletwo
        \def\@oddfoot{\hfil \smash{\Large\sl DRAFT} \hfil}
        \let\@evenfoot\@oddhead}
        \def\@eqnnum{(\theequation)\rlap{\kern\marginparsep\tt\@eqnlabel}%
        \global\let\@eqnlabel\@vacuum}  }

\newcommand{\rf}[1]{(\ref{#1})}
\renewcommand{\theequation}{\thesection.\arabic{equation}}
\renewcommand{\thefootnote}{\fnsymbol{footnote}}
\newcommand{\newsection}{   % Numeration of eqs. is automatic
\setcounter{equation}{0}\section}

\def\appendix#1{\addtocounter{section}{1}\setcounter{equation}{0}
\renewcommand{\thesection}{\Alph{section}}
\section*{Appendix \thesection\protect\indent \parbox[t]{11.15cm}{#1}}
\addcontentsline{toc}{section}{Appendix \thesection\ \ \ #1}}

\def\be{\begin{equation}}
\def\ee{\end{equation}}
\def\beq{\begin{eqnarray}}
\def\eeq{\end{eqnarray}}

\def\parline{\,\partial\kern -0.55em /\,\,}

\def\half{{\frac{1}{2}}}

\def\CC{{\cal C}}

\def\sbf{{\bf s}}

\def\noinbf#1{\noindent {\bf #1}}

\def\phik{|\phi\rangle}
\def\phibr{\langle\phi|}

\def\psik{|\psi\rangle}
\def\psibr{\langle\psi|}

\def\smAdS{{\scriptscriptstyle AdS}}

\def\sm(A)dS{{\scriptscriptstyle (A)dS }}

\def\alphab{\bar\alpha}

\def\irm{{\rm i}}
\def\imrm{{\rm im}}

\def\diff{{\rm diff}}
\def\field{{\rm field}}
\def\bfrm{{\rm bf}}
\def\flat{{\rm flat}}

\def\dsf{{\sf d}}
\def\psf{{\sf p}}
\def\ksf{{\sf k}}
\def\msf{{\sf m}}

\begin{document}

%\draft

\begin{flushright}
FIAN-TD-2021-05  \hspace{1.5cm} \ \ \ \ \ \ \\
arXiv: 2105.11281 [hep-th], V2 \\
\end{flushright}

\vspace{1cm}

\begin{center}

{\Large \bf Mixed-symmetry continuous-spin fields in flat and AdS spaces}

\vspace{2.5cm}

R.R. Metsaev%
\footnote{ E-mail: metsaev@lpi.ru
}

\vspace{1cm}

{\it Department of Theoretical Physics, P.N. Lebedev Physical
Institute, \\ Leninsky prospect 53,  Moscow 119991, Russia }

\vspace{3.5cm}

{\bf Abstract}

\end{center}

In the framework of light-cone gauge approach, bosonic and fermionic mixed-symmetry con\-tinuous-spin fields in AdS space and flat space are considered.
For such fields, the light-cone gauge actions are found. Realization of relativistic symmetries of AdS and flat spaces on the continuous-spin fields is also presented.
Simple realization of spin operators is found. With respect to vector variable entering the continuous-spin fields, the spin operators for the fields in flat space turn out to be first-order differential operators, while, for the the fields in AdS space, the spin operators are realized as the second-order differential operators. By product, we obtain also the simple description of triplectic mixed-symmetry continuous-spin fields.

\vspace{2cm}


\newpage
\renewcommand{\thefootnote}{\arabic{footnote}}
\setcounter{footnote}{0}

%%%%%%%%%%%%%%%%%%%%%%%%%%%%%%%%%%%%%%%%%%%
\newsection{\large Introduction}
%%%%%%%%%%%%%%%%%%%%%%%%%%%%%%

In recent time, continuous-spin field dynamics has attracted some interest. For review, see Refs.\cite{Bekaert:2006py,Bekaert:2017khg,Brink:2002zx}. This is to say that the continuous-spin field dynamics in AdS and flat spaces has actively been studied by various methods. Studies of Lagrangian formulation of massless spin-0 and spin-$\half$ continuous-spin fields in flat space were initiated in Refs.\cite{Schuster:2014hca,Najafizadeh:2015uxa}, while spin-0 and spin-$\half$ continuous-spin fields in AdS space were considered in Refs.\cite{Metsaev:2016lhs}-\cite{Zinoviev:2017rnj}.
Various BRST Lagrangian formulations were investigated in Refs.\cite{Bengtsson:2013vra}-\cite{Buchbinder:2020nxn}.
Interesting discussion of interrelation of continuous-spin field and massive higher-spin field may be found in Refs.\cite{Khan:2004nj}-\cite{Rehren:2017xzn}.
Supersymmetric continuous-spin fields were studied in Refs.\cite{Buchbinder:2019iwi}-\cite{Khabarov:2020glf}.
Problems of interacting continuous-spin fields were considered in Refs.\cite{Metsaev:2017cuz}-\cite{Metsaev:2018moa}. Twistor formulations of continuous-spin fields may be found in Refs.\cite{Buchbinder:2019iwi,Buchbinder:2019sie,Buchbinder:2018soq}.%
\footnote{ Interesting recent discussion of the twistor formulation of superparticle in AdS space may be found in Refs.\cite{Uvarov:2019vmd}.}
Various interesting aspects of the continuous-spin field dynamics were also discussed in Refs.\cite{Rivelles:2014fsa}-\cite{Ponomarev:2010st}.

So far, most of studies in the topic of continuous-spin fields were devoted to Lagrangian formulation of spin-0 and spin-$\half$ continuous-spin fields, while Lagrangian formulation of mixed-symmetry continuous-spin fields was obtained only for some particular cases. Namely, for $AdS_{d+1}$ and $R^{d,1}$ spaces with arbitrary $d$, Lorentz covariant formulation of the bosonic and fermionic mixed-symmetry continuous-spin fields associated with two-rows Young tableaux was obtained in Ref.\cite{Khabarov:2017lth}, while, for $AdS_5$ and $R^{4,1}$ spaces, light-cone gauge Lagrangian formulation of the bosonic mixed-symmetry continuous-spin fields associated with two-rows Young tableaux was obtained in Ref.\cite{Metsaev:2017myp}.%
\footnote{ At the level of equations of motion, BRST formulation of mixed-symmetry and spin-0 continuous-spin fields in flat space may be found in Refs.\cite{Alkalaev:2017hvj} and Ref.\cite{Burdik:2019tzg}. According to the  terminology we use in this paper, mixed-symmetry continuous-spin fields in Refs.\cite{Khabarov:2017lth,Metsaev:2017myp} should be referred to as totally symmetric continuous-spin fields.
}
For $AdS_{d+1}$ and $R^{d,1}$ with arbitrary $d$, Lagrangian formulation of mixed-symmetry continuous-spin fields associated with arbitrary rows Young tableaux has not been studied in the literature. This is what we do in this paper. Namely, for the bosonic and fermionic mixed-symmetry continuous-spin fields propagating in $AdS_{d+1}$ and $R^{d,1}$ spaces with arbitrary $d$ and associated with arbitrary rows Young tableaux, light-cone gauge Lagrangian formulation is obtained in this paper. The remarkable feature of our formulation is that it provides us not only the Lagrangian formulation for the mixed-symmetry continuous-spin fields, but also suggests the simple expressions for spin operators entering the relativistic symmetries of AdS and flat spaces.

\newsection{ \large Massive mixed-symmetry continuous-spin fields in $R^{d,1}$ flat space}

{\bf Field content}. To streamline the light-cone gauge description of bosonic and fermionic mixed-symmetry continuous-spin fields we introduce the ket-vectors  $|\phi(x,z,\zeta,\alpha)\rangle$ and $|\psi(x,z,\zeta,\alpha)\rangle$ respectively.
The arguments $x$, $z$, where $x\equiv x^+,x^-,x^i$, $i=1,\ldots,d-2$, stand for the coordinates of $R^{d,1}$ space,%
\footnote{ We use metric of flat space $ds^2 =-dx_0^2+dx_i^2+dx_{d-1}^2+dz^2$. The coordinates $x^\pm$ are defined as $x^\pm=(x^{d-1} \pm x^0)/\sqrt{2}$ and
$x^+$ is a light-cone time. For the coordinates and derivatives, our conventions  are as follows: $x^I= (x^i, x^d$), $x^d \equiv z$,
$\partial^i=\partial_i\equiv\partial/\partial x^i$,
$\partial_z\equiv\partial/\partial z$, $\partial^\pm=\partial_\mp
\equiv \partial/\partial x^\mp$, where we use the indices
$i,j =1,\ldots, d-2$; $I,J,K,L=1,2,\ldots,d-2, d$. Vectors of the $so(d-1)$
algebra and their scalar product are decomposed as $X^I=(X^i,X^z)$, $X^I Y^I = X^i Y^i + X^zY^z$.}
while the argument $\zeta$ is used to denote a vector of the $so(d-2)$ algebra $\zeta^i$.
The argument $\alpha$ stands for a finite set of the oscillators $\alpha_n^i$,
\be \label{26042021-manus-01-a0}
[\alphab_m^i,\alpha_n^j]= \delta^{ij}\delta_{mn}\,, \quad \alphab_n^i|0\rangle=0\,, \quad \alpha_n^{i\dagger} = \alphab_n^i\,,\quad m,n=1,\ldots,\nu\,, \quad \nu \equiv [\frac{d-2}{2}]\,,\qquad
\ee
where  vector indices of the $so(d-2)$ algebra take values $i,j=1,\ldots,d-2$. Note that the commonly used light-cone gauge approach for discrete-spin fields in $R^{d,1}$ is realized by using the manifest $so(d-1)$ algebra symmetries. However, as has already been noted in the literature, (see, e.g., Refs.\cite{Bekaert:2006py,Brink:2002zx}), continuous-spin field in $R^{d,1}$ is described by the irrep of the $so(d-2)$ algebra. For this reason, it is the light-cone approach based on the realization of the manifest $so(d-2)$ algebra symmetries that turns out to be very convenient for us in this paper.

Infinite number of ordinary light-cone gauge continuous-spin fields depending on the space-time coordinates $x^\pm$, $x^i$, $z$ and the variable $\zeta^i$ are obtained by expanding the ket-vectors into the oscillators $\alpha_n^i$. Power expansions of the ket-vectors $\phik$, $\psik$ in the oscillators are similar,
\beq
\label{26042021-manus-01-a1} && |\phi(x,z,\zeta,\alpha)\rangle = \sum_{s_1,\ldots,
s_\nu = 0}^{\infty}\,\,|\phi_{s_1 \ldots s_\nu }(x,z,\zeta,\alpha)\rangle\,,
\\
\label{26042021-manus-01-a2} && |\phi_{s_1 \ldots s_\nu }
(x,z,\zeta,\alpha)\rangle \equiv \prod_{n=1}^\nu
\alpha_n^{i_1^n} \ldots \alpha_n^{i_{s_n}^n}\,
\phi_{s_1\ldots s_\nu}^{i_1^1\ldots i_{s_1}^1\ldots i_1^\nu \ldots
i_{s_\nu }^\nu }(x,z,\zeta)|0\rangle\,,
\\
\label{26042021-manus-01-a1b} && |\psi(x,z,\zeta,\alpha)\rangle = \sum_{s_1,\ldots,
s_\nu = 0}^{\infty}\,\,|\psi_{s_1 \ldots s_\nu }(x,z,\zeta,\alpha)\rangle\,,
\\
\label{26042021-manus-01-a2b} && |\psi_{s_1 \ldots s_\nu }
(x,z,\zeta,\alpha)\rangle \equiv \prod_{n=1}^\nu
\alpha_n^{i_1^n} \ldots \alpha_n^{i_{s_n}^n}\,
\psi_{s_1\ldots s_\nu}^{i_1^1\ldots i_{s_1}^1\ldots i_1^\nu \ldots
i_{s_\nu }^\nu }(x,z,\zeta)|0\rangle\,,
\eeq
where $s_n$ are integers. Spinor indices of the tensor-spinor fermionic fields in \rf{26042021-manus-01-a2b}  are implicit.
We see that the ket-vectors in \rf{26042021-manus-01-a2} and \rf{26042021-manus-01-a2b} are degree-$s_n$ homogeneous polynomials in $\alpha_n^i$,
\be \label{26042021-manus-01-a3}
(\alpha_n^i \alphab_n^i - s_n) |\phi_{s_1 \ldots s_\nu }\rangle = 0 \,, \hspace{1cm}   (\alpha_n^i \alphab_n^i - s_n) |\psi_{s_1 \ldots s_\nu }\rangle = 0 \,, \hspace{1cm} \hbox{for all}\,\, n\,.
\ee
In $R^{d,1}$, physical D.o.F of the continuous-spin field are described by irrep of the $so(d-2)$ algebra, which can be labeled by $\sbf^\bfrm$, where
\beq
\label{26042021-manus-01-a5} && \sbf^\bfrm = s_1^\bfrm,\,s_2^\bfrm,\,\ldots,\, s_\nu^\bfrm\,, %
\\
\label{26042021-manus-01-a6} && s_n^\bfrm = s_n\,,  \hspace{0.4cm} \hbox{for bosonic field};
\hspace{1cm} s_n^\bfrm = s_n+\half\,,    \hspace{0.4cm} \hbox{for fermionic field};
\\
\label{26042021-manus-01-a6x} && s_1\geq s_2 \geq \ldots \geq s_{\nu-1}\geq s_\nu \geq 0\,.
\eeq
For odd $d$, the $\sbf^\bfrm$ is the highest weight of irrep of the $so(d-2)$ algebra.%
\footnote{ For even $d$, there are two highest weights with $\pm s_\nu$ for bosonic fields and $\pm (s_\nu+\half)$ for fermionic fields.}
The irrep with $\sbf^\bfrm$ as in \rf{26042021-manus-01-a5} can be described by tensor (tensor-spinor) field whose $so(d-2)$ space tensor indices have the structure of the Young tableaux. For such Young tableaux, $s_n$ is equal to length of the $n$-th row. Let us now explain our terminology. If $s_n=0$ for all $n$, then  fields $|\phi_{s_1 \ldots s_\nu }\rangle$ \rf{26042021-manus-01-a2} and $|\psi_{s_1 \ldots s_\nu }\rangle$ \rf{26042021-manus-01-a2b} are referred to as spin-0 and spin-$\half$ continuous-spin fields respectively. If $s_1>0$, $s_n=0$ for $n=2,\ldots,\nu$, then  fields $|\phi_{s_1 \ldots s_\nu }\rangle$ \rf{26042021-manus-01-a2} and $|\psi_{s_1 \ldots s_\nu }\rangle$ \rf{26042021-manus-01-a2b} are referred to totally symmetric continuous-spin fields. If $s_1\geq 2$, $s_2\geq 1$, then  $|\phi_{s_1 \ldots s_\nu }\rangle$ \rf{26042021-manus-01-a2} and $|\psi_{s_1 \ldots s_\nu }\rangle$ \rf{26042021-manus-01-a2b} are referred to as mixed-symmetry continuous-spin fields.

As is well known for the ket-vectors $|\phi_{s_1 \ldots s_\nu }\rangle$ \rf{26042021-manus-01-a2} and $|\psi_{s_1 \ldots s_\nu }\rangle$ \rf{26042021-manus-01-a2b} to be carriers of the $so(d-2)$ algebra irreps we should impose the following constraints:%
\footnote{ For bosonic field, even $d$, and $s_\nu\ne0$, $\nu=\frac{d-2}{2}$, constraints   \rf{26042021-manus-01-a8},\rf{26042021-manus-01-a9} should be supplemented by self-duality constraint. For fermionic field and even $d$, constraints   \rf{26042021-manus-01-a8}-\rf{26042021-manus-01-a12} should be supplemented by the chirality constraint.}
\beq
\label{26042021-manus-01-a8} && \alpha_m^i \alphab_n^i|\phi_{s_1 \ldots s_\nu }\rangle = 0 \,,  \hspace{1cm} \alpha_m^i \alphab_n^i|\psi_{s_1 \ldots s_\nu }\rangle = 0 \,, \hspace{2cm} \hbox{for}\,\,\, m < n \,; \hspace{2cm}
\\
\label{26042021-manus-01-a9} && \alphab_m^i \alphab_n^i |\phi_{s_1 \ldots s_\nu }\rangle = 0 \,, \hspace{1cm} \alphab_m^i \alphab_n^i |\psi_{s_1 \ldots s_\nu }\rangle = 0 \,, \hspace{2cm} \hbox{for all}\,\,\,  m,n\,; \hspace{2cm}
\\
\label{26042021-manus-01-a12} && \hspace{4.4cm} \gamma^i \alphab_n^i|\psi_{s_1 \ldots s_\nu }\rangle = 0 \,,  \hspace{2.2cm} \hbox{for all}\,\, n\,,
\eeq
where $\gamma^i$ stands for the Dirac gamma matrices of the $so(d-2)$ algebra.
To summarize it is the ket-vectors $|\phi_{s_1 \ldots s_\nu }\rangle$ \rf{26042021-manus-01-a2} and $|\psi_{s_1 \ldots s_\nu }\rangle$ \rf{26042021-manus-01-a2b} subject to constraints \rf{26042021-manus-01-a3}-\rf{26042021-manus-01-a12} that describe irreducible bosonic and fermionic mixed-symmetry continuous-spin fields in $R^{d,1}$. In what follows, in order to treat arbitrary spin fields on an equal footing we collect ket-vectors $|\phi_{s_1 \ldots s_\nu }\rangle$ \rf{26042021-manus-01-a2} and $|\psi_{s_1 \ldots s_\nu }\rangle$ \rf{26042021-manus-01-a2b} into the respective  ket-vectors $\phik$ \rf{26042021-manus-01-a1} and $\psik$ \rf{26042021-manus-01-a1b} and present our results entirely in terms of the $\phik$ and $\psik$. We note that the bosonic and fermionic continuous-spin fields are {\it complex-valued}.

{\bf Light-cone gauge action}. Light-cone gauge actions take the form:%
\beq
\label{26042021-manus-01} &&  S  = \int d^{d+1}x\, d^{d-2}\zeta\langle \phi|\bigl(\Box + \partial_z^2 - m^2 \bigr)|\phi\rangle\,, \hspace{1.7cm} \hbox{for bosonic field},
\\
\label{26042021-manus-02} && S  = \int  d^{d+1}x\,  d^{d-2}\zeta \langle \psi|\frac{\irm }{\partial^+}\bigl(\Box + \partial_z^2 - m^2\bigr)|\psi\rangle\,, \hspace{1cm} \hbox{for fermionic field},
\\
&& \hspace{2cm} \Box \equiv 2\partial^+\partial^- + \partial^i\partial^i\,, \qquad d^{d+1} x \equiv dx^+ dx^-dz d^{d-2} x \,,
\eeq
where $\phibr = (\phik)^\dagger$, $\psibr = (\psik)^\dagger$. Note that, for a massive continuous-spin fields, $m^2< 0$.

The light-cone gauge spoils the Poincar\'e symmetries of fields in $R^{d,1}$. In order to show that the Poincar\'e symmetries are still present we have to find Noether charges which generate them. For free fields, Noether charges (generators)
take the following form in terms of the $\phik$ and $\psik$:
\beq
\label{26042021-manus-03}  &&  G_\field = 2 \int dz dx^- d^{d-2}x \, d^{d-2}\zeta\, \langle\partial^+\phi|G_\diff|\phi\rangle\,, \hspace{1.6cm} \hbox{for bosonic field}, \hspace{2cm}
\\
\label{26042021-manus-04} && G_\field = 2 \int dz dx^- d^{d-2} x \, d^{d-2}\zeta\, \langle \psi|G_\diff|\psi\rangle\,, \hspace{2cm} \hbox{for fermionic field}, \hspace{2cm}
\eeq
where $G_\diff$ are differential operators acting on the ket-vectors. We then note that actions \rf{26042021-manus-01} and \rf{26042021-manus-02}  are invariant under the respective transformations $\delta\phik =G_\diff\phik$ and $\delta\psik =G_\diff\psik$. The operators $G_\diff$ take the following form:%
\footnote{In  basis of the $so(d-1)$ algebra, generators of the Poincar\'e algebra $iso(d,1)$ are spanned by $P^\pm,P^I$, $J^{IJ},J^{\pm I}$. In basis of the $so(d-2)$ algebra, the $P^I$, $J^{IJ}$, $J^{\pm I}$ are decomposed as $P^I=P^i,P^z$, $J^{IJ}=J^{ij},J^{zi}$, $J^{\pm I}=J^{\pm i},J^{\pm z}.$
}
\beq
\label{26042021-manus-05} && P^i=\partial^i\,, \qquad P^z=\partial_z\,, \qquad  P^+=\partial^+\,, \hspace{1cm} P^-=\frac{m^2-\partial^i\partial^i-\partial_z\partial_z}{2\partial^+}\,,
\\
\label{26042021-manus-06} && J^{+-} = x^+ P^- -x_\bfrm^-\partial^+\,,
\\
\label{26042021-manus-07} && J^{ij} = x^i\partial^j-x^j\partial^i + M^{ij}\,, \hspace{1.4cm} J^{zi} = z\partial^i - x^i\partial_z + M^{zi}\,,
\\
\label{26042021-manus-08} && J^{+i} = x^+\partial^i - x^i\partial^+\,, \hspace{2.4cm} J^{+z}=x^+\partial_z -z\partial^+\,,
\\
\label{26042021-manus-09} && J^{-i}=x_\bfrm^- \partial^i - x^i P^- +M^{-i}\,, \hspace{1cm} J^{-z} = x_\bfrm^-\partial_z - z P^- +M^{-z}\,,
\\
\label{26042021-manus-10} && \hspace{1cm} M^{-i}  =  M^{ij} \frac{\partial^j}{\partial^+} - M^{zi} \frac{\partial_z}{\partial^+} + \frac{1}{\partial^+}  M^i\,,
\\
\label{26042021-manus-11} && \hspace{1cm} M^{-z}  =  M^{zi} \frac{\partial^i}{\partial^+} + \frac{1}{\partial^+}  M^z\,.
\eeq
We now explain the notation we use in \rf{26042021-manus-05}-\rf{26042021-manus-11}. Symbol $x_\bfrm^-$ appearing in \rf{26042021-manus-06},\rf{26042021-manus-09} is defined in terms of the coordinate $x^-$ and the derivative $\partial^+=\partial/\partial x^-$ as
\be \label{26042021-manus-12}
x_{\bfrm}^- = x^- \,,  \hspace{0.7cm} \hbox{for bosonic field}; \hspace{1.5cm}  x_{\bfrm}^- = x^- + \frac{1}{2\partial^+}\,, \hspace{0.7cm} \hbox{for fermionic field}.\quad
\ee
The spin operators $M^{ij}$, $M^{zi}$, $M^i$, $M^z$ depend only on the variable $\zeta^i$, the derivatives $\partial/\partial \zeta^i$, and the oscillators.
Operators $M^{IJ}=M^{ij},M^{zi}$ and $M^I=M^i,M^z$ satisfy the commutators and the hermicity conditions given by
\beq
\label{26042021-manus-14} && [M^{IJ}, M^{KL}] =\delta^{JK} M^{IL} + 3 \hbox{ terms}, \qquad [M^I,M^{JK}] =\delta^{IJ} M^K - \delta^{IK} M^J\,,
\\
\label{26042021-manus-15} && [M^I,M^J] = m^2  M^{IJ} \,,
\\
\label{26042021-manus-16} && \hspace{3cm}  M^{IJ\dagger} = - M^{IJ}\,, \qquad M^{I\dagger} = M^I\,,
\eeq
where $\delta^{IJ}=\delta^{ij},\delta^{zz}$, $\delta^{zz}=1$. As seen from \rf{26042021-manus-14}, the $M^{IJ}$ is spin  operator of the $so(d-1)$ algebra, while the $M^I$ transforms as vector under transformations of the $so(d-1)$ algebra.

Commutators \rf{26042021-manus-15}  are basic equations of the light-cone gauge formalism in the flat space. Our main result in this Section is the solution for the spin operators $M^{IJ}$, $M^I$ given by
\beq
\label{26042021-manus-17} && M^{zi} = \half \psf^i + \ksf^i\,,
\\
\label{26042021-manus-18} &&  M^{ij} = \msf^{ij}\,,
\\
\label{26042021-manus-19} && M^z =   m_\imrm \dsf \,,
\\
\label{26042021-manus-20} && M^i = m_\imrm \bigl( - \half \psf^i + \ksf^i\bigr)\,,
\\
\label{26042021-manus-21} && m_\imrm \equiv \irm \sqrt{-m^2}\,, \qquad m^2 < 0\,,\qquad m_\imrm^* = - m_\imrm\,,
\eeq
where operators $\psf^i$, $\ksf^i$, $\dsf$,  $\msf^{ij}$ are defined as follows
\beq
\label{26042021-manus-22} && \psf^i = \partial_{\zeta^i}\,, \qquad  \partial_{\zeta^i}\equiv  \partial/\partial \zeta^i\,,
\\
\label{26042021-manus-23} && \ksf^i = -\half \zeta^j\zeta^j \partial_{\zeta^i} + \zeta^i \dsf + m^{ij} \zeta^j\,,
\\
\label{26042021-manus-24} && \dsf = \zeta^i\partial_{\zeta^i} + \frac{d-2}{2} + s_\imrm \,, \qquad s_\imrm ^* =  - s_\imrm \,,
\\
\label{26042021-manus-25} && \msf^{ij} = \zeta^i \partial_{\zeta^j} - \zeta^j \partial_{\zeta^i} + m^{ij}\,.
\eeq
Parameters $m_\imrm$ \rf{26042021-manus-21} and $s_\imrm$ \rf{26042021-manus-24} are purely imaginary numbers. In \rf{26042021-manus-23},\rf{26042021-manus-25} and below, operator $m^{ij}$ stands for spin operator of the $so(d-2)$ algebra.  This operator depends only on the oscillators and satisfies the commutation relations
\be
[m^{ij},m^{kl}] = \delta^{jk} m^{il} + 3 \hbox{ terms}.
\ee
Realization of the $m^{ij}$ on space of the ket-vectors $\phik$ and $\psik$ takes the form
\beq
&& m^{ij} = \sum_{n=1}^\nu m_n^{ij}\,, \hspace{2.3cm} \hbox{for bosonic field} \,\,\, \phik;
\\
&& m^{ij} = \half \gamma^{ij} + \sum_{n=1}^\nu m_n^{ij}\,,   \hspace{1cm} \hbox{for fermionic field}\,\,\, \psik;
\\
&& \hspace{1cm} m_n^{ij} \equiv \alpha_n^i\alphab_n^j - \alpha_n^j\alphab_n^i \,, \hspace{1cm} \gamma^{ij} \equiv \half  \gamma^i\gamma^j - (i \leftrightarrow j)\,,
\eeq
where $\gamma^i$ stands for the Dirac gamma matrices of the $so(d-2)$ algebra.

The operators given in \rf{26042021-manus-22}-\rf{26042021-manus-25} satisfy the following commutators of the $so(d-2,2)$ algebra and the hermitian conjugation rules
\beq
\label{26042021-manus-26} && [\dsf,\psf^i] = - \psf^i\,, \qquad [\dsf,\ksf^i] = \ksf^i\,,
\qquad
[{\sf p}^i,{\sf p}^j]=0\,,
\qquad
[{\sf k}^i,{\sf k}^j]=0\,,
\\
\label{26042021-manus-27} && [\psf^i,\msf^{jk}] = \delta^{ij} \psf^k  - \delta^{ik} \psf^j\,,
\hspace{1.8cm}
[\ksf^i,\msf^{jk}] = \delta^{ij} \ksf^k - \delta^{ik} \ksf^j\,,
\\
\label{26042021-manus-28} && [\psf^i,\ksf^j] = \delta^{ij} \dsf - \msf^{ij}\,, \hspace{2.5cm} [\msf^{ij},\msf^{kl}]
= \delta^{jk} \msf^{il} + 3 \hbox{ terms},
\\
\label{26042021-manus-30} && \psf^{i\dagger} = - \psf^i\,, \qquad \ksf^{i\dagger} = - \ksf^i\,, \qquad \dsf^\dagger = - \dsf\,,  \qquad \msf^{ij\dagger} = - \msf^{ij}\,,
\eeq
where the rules in \rf{26042021-manus-30} are simply obtained from the following hermitian conjugation rules:
\be \label{26042021-manus-29}
\zeta^{i\dagger} = \zeta^i\,, \qquad \partial_{\zeta^i}^\dagger = - \partial_{\zeta^i}\,,\qquad m^{ij\dagger} = - m^{ij}\,.
\ee
In turn, relations \rf{26042021-manus-30} and \rf{26042021-manus-17}-\rf{26042021-manus-20} imply the hermicity conjugation rules given in \rf{26042021-manus-16}.

We finish this section by the following remarks.

\noinbf{i}) The massive mixed-symmetry continuous-spin field in $R^{d,1}$ is labeled by the parameters $m_\imrm$, $s_\imrm$, and $\sbf^\bfrm$. The $m_\imrm$, $s_\imrm$ are purely imaginary parameters, where $m_\imrm$ is expressed in terms of the square of mass  \rf{26042021-manus-21}. The $\sbf^\bfrm$ defined in \rf{26042021-manus-01-a3}-\rf{26042021-manus-01-a6x} labels irrep of the $so(d-2)$ algebra.

\noinbf{ii}) The spin operators \rf{26042021-manus-17}-\rf{26042021-manus-20} turn out to be polynomial in the variable $\zeta^i$ and degree-1 polynomials in the derivative $\partial/\partial \zeta^i$. This very important feature of the spin operators will make easier future study of interacting continuous-spin fields.

\noinbf{iii}) To deal with the irreducible continuous-spin fields we should use the ket-vectors subject to constraints in \rf{26042021-manus-01-a3}-\rf{26042021-manus-01-a12}. If we ignore those constraints, then our ket-vectors $\phik$ and $\psik$ describe so called triplectic continuous-spin fields. Thus our formulation, among other things, provides easy and quick access to the
triplectic mixed-symmetry continuous-spin fields.

\newsection{ \large Mixed-symmetry continuous-spin fields in $AdS_{d+1}$ space}

For  AdS space, the light-cone gauge approach was developed in Refs.\cite{Metsaev:1999ui,Metsaev:2003cu}. In this section we adapt the formulation obtained in Ref.\cite{Metsaev:2003cu} for the study of bosonic and fermionic continuous-spin fields  and present our results.

{\bf Field content}. To discuss light-cone gauge mixed-symmetry bosonic and fermionic continuous-spin fields we introduce the respective ket-vectors $|\phi(x,z,\zeta,\alpha)\rangle$ and $|\psi(x,z,\zeta,\alpha)\rangle$. The arguments $x$, $z$, where $x\equiv x^+,x^-,x^i$, $i=1,\ldots,d-2$, stand for the coordinates of $AdS_{d+1}$ space,%
\footnote{ We use metric of $AdS_{d+1}$, $ds^2=R^2(-dx_0^2+dx_i^2+dx_{d-1}^2+dz^2)/z^2$. The coordinates $x^\pm$ are defined as $x^\pm=(x^{d-1} \pm x^0)/\sqrt{2}$ and
$x^+$ is a light-cone time. For coordinates and derivatives, our conventions are as follows: $x^I= (x^i, x^d$), $x^d \equiv z$,
$\partial^i=\partial_i\equiv\partial/\partial x^i$,
$\partial_z\equiv\partial/\partial z$, $\partial^\pm=\partial_\mp
\equiv \partial/\partial x^\mp$, where we use the indices
$i,j =1,\ldots, d-2$; $I,J,K,L=1,2,\ldots,d-2, d$. Vectors of the $so(d-1)$
algebra and their scalar product are decomposed as $X^I=(X^i,X^z)$, $X^I Y^I = X^i Y^i + X^zY^z$. We use the shortcut $x^Ix^I=x^ix^i+z^2$.}
while the argument $\zeta$ stands for a vector of the $so(d-2)$ algebra $\zeta^i$.
The argument $\alpha$ stands for a finite set of the oscillators $\alpha_n^i$ given in \rf{26042021-manus-01-a0}, where vector indices of the $so(d-2)$ algebra take values $i,j=1,\ldots,d-2$.
Infinite number of ordinary light-cone gauge continuous-spin fields depending on the space-time coordinates $x^\pm$, $x^i$, $z$ and the variable $\zeta^i$ are obtained by expanding the ket-vectors $|\phi(x,z,\zeta,\alpha)\rangle$, $|\psi(x,z,\zeta,\alpha)\rangle$ into the oscillators $\alpha_n^i$ as in \rf{26042021-manus-01-a1}-\rf{26042021-manus-01-a2b}.  In order to deal with irreducible continuous-spin fields in $AdS_{d+1}$, the ket-vectors $|\phi_{s_1 \ldots s_\nu}\rangle$ and $|\psi_{s_1 \ldots s_\nu }\rangle$ should satisfy the respective constraints in \rf{26042021-manus-01-a3}-\rf{26042021-manus-01-a12}. Ignoring those constraints, we get so called triplectic continuous-spin fields.
The bosonic and fermionic continuous-spin AdS fields are {\it complex-valued}.

{\bf Light-cone gauge action}. Light-cone gauge actions for the bosonic and fermionic continuous-spin AdS fields take the form:
\beq
\label{27042021-manus-01} && S  = \int d^{d+1}x d^{d-2}\zeta\,\langle \phi|\bigl(\Box + \partial_z^2 - \frac{1}{z^2}A\bigr)|\phi\rangle\,, \hspace{1.7cm} \hbox{for bosonic field},
\\
\label{27042021-manus-02}  && S  = \int d^{d+1}x d^{d-2}\zeta\, \langle \psi|\frac{\irm }{\partial^+}\bigl(\Box + \partial_z^2 - \frac{1}{z^2}A\bigr)|\psi\rangle\,, \hspace{1cm} \hbox{for fermionic field},
\\
\label{27042021-manus-03}  && \hspace{1cm} \Box = 2\partial^+\partial^- + \partial^i\partial^i \,,\qquad d^{d+1}x =dz dx^+dx^- d^{d-2}x\,,
\eeq
$\phibr = (\phik)^\dagger$, $\psibr = (\psik)^\dagger$. Operator $A$ depends only on the variable $\zeta^i$ and the derivative $\partial/\partial \zeta^i$.

Relativistic symmetries of fields in $AdS_{d+1}$ are described by the $so(d,2)$ algebra.  The light-cone gauge spoils the $so(d,2)$ algebra symmetries. Therefore in order to show that the $so(d,2)$ algebra symmetries are maintained in our approach we should find the
Noether charges which generate them. For free bosonic and fermionic fields, Noether charges (generators) can be presented in the following respective forms:
\beq
\label{27042021-manus-04} &&    G_\field = 2 \int dz dx^- d^{d-2} x \, d^{d-2}\zeta\, \langle\partial^+\phi|G_\diff|\phi\rangle\,, \hspace{1.6cm} \hbox{for bosonic field},
\\
\label{27042021-manus-05} && G_\field = 2 \int dz dx^- d^{d-2} x\, d^{d-2}\zeta\, \langle \psi|G_\diff|\psi\rangle\,, \hspace{2cm} \hbox{for fermionic field},
\eeq
where $G_\diff$ are differential operators acting on the ket-vectors.  Light-cone gauge actions \rf{27042021-manus-01} and \rf{27042021-manus-02} are invariant under the respective transformations $\delta\phik =G_\diff\phik$ and $\delta\psik =G_\diff\psik$. The  operators $G_\diff$ take the following form:
\beq
\label{27042021-manus-06} && P^i=\partial^i\,, \qquad  P^+=\partial^+\,, \hspace{1cm} P^-=-\frac{\partial^i\partial^i + \partial_z\partial_z}{2\partial^+} +\frac{1}{2z^2\partial^+}A\,,
\\
\label{27042021-manus-07} && J^{+-} = x^+ P^- -x_\bfrm^-\partial^+\,, \hspace{1.3cm} J^{ij} = x^i\partial^j-x^j\partial^i + M^{ij}\,,
\\
\label{27042021-manus-08} && J^{+i}=x^+\partial^i-x^i\partial^+\,, \hspace{1.9cm} J^{-i} = x_\bfrm^-\partial^i-x^i P^- +M^{-i}\,,
\\
\label{27042021-manus-09} && D = x^+ P^- +x_\bfrm^-\partial^+ + x^i\partial^i + z\partial_z + \frac{d-1}{2}\,,
\\
\label{27042021-manus-10} && K^+ = -\frac{1}{2}(2x^+x_\bfrm^- + x^Jx^J)\partial^+ + x^+D\,,
\\
\label{27042021-manus-11} && K^i = -\frac{1}{2}(2x^+ x_\bfrm^- + x^Jx^J)\partial^i +x^i D+M^{iJ}x^J+M^{i-}x^+\,,
\\
\label{27042021-manus-12} && K^-=-\frac{1}{2}(2x^+ x_\bfrm^- + x^J x^J) P^- + x_\bfrm^- D + \frac{1}{\partial^+}x^I\partial^JM^{IJ}
- \frac{x^i}{2z\partial^+}[M^{zi},A] +\frac{1}{\partial^+}B\,,\qquad
\\
\label{27042021-manus-13} && \hspace{0.8cm} M^{-i} \equiv M^{iJ}\frac{\partial^J}{\partial^+}
-\frac{1}{2z\partial^+}[M^{zi},A]\,,\qquad M^{-i}=-M^{i-}\,,\quad M^{IJ}=- M^{JI}\,.
\eeq
We now explain notation we use in \rf{27042021-manus-06}-\rf{27042021-manus-13}. Symbol $x_\bfrm^-$ is defined as in \rf{26042021-manus-12}. Operators $A$, $B$, $B^I$ $M^{IJ}$ appearing in \rf{27042021-manus-06}-\rf{27042021-manus-13} depend only on the variable $\zeta^i$, the derivative $\partial/\partial \zeta^i$ and the oscillators.%
\footnote{ Commutators of the generators \rf{27042021-manus-06}-\rf{27042021-manus-13} we use in this paper may be found in (A6),(A7) in Ref.\cite{Metsaev:2015rda}.
}
Note also the decompositions $M^{IJ} = M^{zi}, M^{ij}$ and $B^I = B^z, B^i$. Also note that the operators $M^{IJ}$ and $B^I$ constitute a base of spin operators. The operators $A$ and $B$ can be expressed in terms of the operators $M^{IJ}$ and $B^z$ as
\beq
\label{27042021-manus-14} A & = & \CC_2 + 2B^z + 2M^{zi}M^{zi}+\frac{1}{2}M^{ij}M^{ij} +\frac{d^2-1}{4}\,,
\\
\label{27042021-manus-15} B & = & B^z + M^{zi}M^{zi}\,,
\eeq
where $\CC_2$ is an eigenvalue of the 2nd-order Casimir operator of the $so(d,2)$ algebra (for our conventions on the $\CC_2$, see Appendix A in Ref.\cite{Metsaev:2019opn}). The spin operators $M^{IJ}$, $B^I$  obey the following commutation relations and hermitian conjugation rules
\beq
\label{27042021-manus-16} && [M^{IJ}, M^{KL}] =\delta^{JK} M^{IL} + 3 \hbox{ terms}\,, \qquad [B^I,M^{JK}] =\delta^{IJ}B^K - \delta^{IK}B^J\,,
\\
\label{27042021-manus-17} && [B^I,B^J] = \Bigl( \CC_2 + \frac{1}{2} M^2 + \frac{d^2-3d+4}{2} \Bigr)M^{IJ} - (M^3)^{[I|J]}\,,
\\
\label{27042021-manus-19} &&   M^2 \equiv M^{IJ} M^{IJ}\,, \qquad (M^3)^{[I|J]} \equiv \half M^{IK}M^{KL}M^{LJ}-  (I\leftrightarrow J)\,,
\\
\label{27042021-manus-20} &&  M^{IJ\dagger} = - M^{IJ}\,, \qquad B^{I\dagger} = B^I\,,
\eeq
where $\delta^{IJ}=\delta^{ij},\delta^{zz}$, $\delta^{zz}=1$. The operator $M^{IJ}$ is a spin operator of the $so(d-1)$ algebra, while the operator $B^I$ transforms as vector under transformations of the $so(d-1)$ algebra.  The commutators given in \rf{27042021-manus-17} are basic equations of our light-cone gauge formulation of field dynamics in $AdS_{d+1}$ space. The solution for the operators $M^{IJ}$, $B^I$ we find is given by
\beq
\label{27042021-manus-21} &&  M^{zi} = \half \psf^i + \ksf^i\,,
\\
\label{27042021-manus-22} && M^{ij} = \msf^{ij}\,,
\\
\label{27042021-manus-23} && B^z = e_\imrm\, \dsf + \frac{1}{4}\psf^i\psf^i - \ksf^i \ksf^i\,,\qquad e_\imrm^* = - e_\imrm\,,
\\
\label{27042021-manus-24} && B^i = e_\imrm  \big( - \half \psf^i + \ksf^i\big) +  \half \{\msf^{ij},\half \psf^j - \ksf^j\} + \half \{\dsf,\half \psf^i + \ksf^i\}
\,,
\eeq
where operators $\psf^i$, $\ksf^i$, $\dsf$,  $\msf^{ij}$ take the same form as in \rf{26042021-manus-22}-\rf{26042021-manus-25} and we use the notation $\{x,y\}\equiv xy+yx$.  The parameters $e_\imrm$ in \rf{27042021-manus-23},\rf{27042021-manus-24} and $s_\imrm$ in \rf{26042021-manus-24} are purely imaginary numbers. In terms of the $e_\imrm$, $s_\imrm$, and the spin operator $m^{ij}$, the 2nd-order Casimir operator of the $so(d,2)$ algebra takes the form
\be \label{27042021-manus-25}
\CC_2  = e_\imrm ^2 + s_\imrm ^2 - \frac{d^2+(d-2)^2}{4} + \CC_{so(d-2),\, 2}\,, \qquad \CC_{so(d-2),\,2} = - \half m^{ij} m^{ij}\,,
\ee
where $\CC_{so(d-2),\,2}$ is the 2nd-order Casimir operator of the $so(d-2)$ algebra. We recall that, for the irrep of the $so(d-2)$ algebra with labels as in \rf{26042021-manus-01-a5}, the  $\CC_{so(d-2),\,2}$ is diagonalized as
\be \label{27042021-manus-25-a1}
\CC_{so(d-2),\,2} = \sum_{n=1}^\nu s_n^\bfrm(s_n^\bfrm + d-2 -2n)\,.
\ee
We note that, for the irreducible continuous-spin field, we should use $\CC_2$ \rf{27042021-manus-25} with the $\CC_{so(d-2),\,2}$ given in \rf{27042021-manus-25-a1}, while, for the triplectic continuous-spin field, we should use $\CC_2$ \rf{27042021-manus-25} with $\CC_{so(d-2)\,2}$  given in the 2nd relation in \rf{27042021-manus-25}.

Using relations \rf{27042021-manus-14},\rf{27042021-manus-15}, we get the following representation for the operators $A$ and $B$:
\beq
\label{27042021-manus-26} && A =\psf^i \psf^i +  ( \dsf + e_\imrm   )^2 - \frac{1}{4}\,,
\\
\label{27042021-manus-27} && B =   e_\imrm  \, \dsf + \half \psf^i\psf^i + \half \{ \psf^i,\ksf^i\}  \,.
\eeq

As an side remark we note that the operator $B^I=B^z,B^i$ \rf{27042021-manus-23},\rf{27042021-manus-24} can be presented in the following $so(d-1)$ covariant form:
\beq
\label{27042021-manus-28} && B^I = e_\imrm  L^I - \half \{M^{IJ},L^J\}\,,
\\
\label{27042021-manus-29} && \hspace{1cm} L^I = L^z,L^i\,, \qquad L^z  \equiv \dsf\,, \hspace{1cm} L^i \equiv  - \half \psf^i + \ksf^i\,,
\eeq
where we introduce new operator $L^I$  defined in \rf{27042021-manus-29}, while the operators $\psf^i$, $\ksf^i$, $\dsf$ take the same form as in \rf{26042021-manus-22}-\rf{26042021-manus-24}.
One can verify that the operators $L^I$, $M^{IJ}$ satisfy the commutators
\be  \label{27042021-manus-31}
[L^I,L^J]= M^{IJ}\,,\qquad [L^I,M^{JK}] =\delta^{IJ} L^K - \delta^{IK} L^J\,.
\ee
Note that $L^{I\,\dagger}=-L^I$. The 2nd commutator in \rf{27042021-manus-31} tells us that the operator $L^I$ transforms as a vector under transformations of the $so(d-1)$ algebra.  Commutators in \rf{27042021-manus-16},\rf{27042021-manus-31} tell us then that the operators $M^{IJ}$ and $L^I$ form the $so(d-1,1)$ algebra. Representation for the $B^I$ in \rf{27042021-manus-28} turns out to be convenient to verify that the $B^I$ satisfies the basic equations \rf{27042021-manus-17}.

{\bf Flat space limit}. In the flat space limit, our continuous-spin AdS fields are realized as the  massive continuous-spin fields studied in the previous section. To demonstrate this fact, let us use the notation $z_\smAdS$ and $z_\flat$ for the respective $z$ coordinate in AdS space in this section and $z$ coordinate in flat space in the previous section. To realize the flat space limit we use the relation
\be \label{27042021-manus-32}
z_\smAdS = R e^{z_\flat /R}\,,
\ee
where $R$ is  radius of AdS space, and take
\be \label{27042021-manus-33}
R \, \longrightarrow \,  \infty\,, \qquad e_\imrm \big|_{R\rightarrow \infty} \, \longrightarrow \,   R\, m_\imrm\,, \qquad z_\flat\,,\, x^\pm,\, x^i,\, \zeta^i,\, s_\imrm - \hbox{ fixed},
\ee
where $m_\imrm$ \rf{26042021-manus-21} is the mass parameter  of the continuous-spin field in the flat space. For large $R$, we get then the relations
\be \label{27042021-manus-34}
\CC_2\big|_{R\rightarrow \infty} \, \longrightarrow \,  R^2 m^2\,, \qquad
B^I\big|_{R\rightarrow \infty} \, \longrightarrow \,  RM^I\,, \qquad
\ee
where $M^I$ is the spin operator in the flat space \rf{26042021-manus-19},\rf{26042021-manus-20}.
Now, using \rf{27042021-manus-33},\rf{27042021-manus-34}, we note that generators of the $so(d,2)$ algebra, denoted as $G_\smAdS$, are related to generators of the Poincar\'e algebra $iso(d,1)$, denoted as $G_\flat$, in the following way:
\beq
\label{27042021-manus-35} && P_\smAdS^a\big|_{R\rightarrow \infty} \longrightarrow P_\flat^a\,,\hspace{1.5cm}  J_\smAdS^{ab}\big|_{R\rightarrow \infty} \longrightarrow J_\flat^{ab}\,,
\\
\label{27042021-manus-36} &&  D_\smAdS\big|_{R\rightarrow \infty} \longrightarrow R P_\flat^z\,, \hspace{1cm} K_\smAdS^a\big|_{R\rightarrow \infty} \longrightarrow - \half R^2 P_\flat^a - R J_\flat^{za}\,,
 \qquad
\eeq
where the $so(d-1,1)$ algebra vector indices $a,b$ take values $+,-,i$, \ \ $i=1,\ldots,d-2$.

We now finish this section by the following remarks.

\noinbf{i}) The mixed-symmetry continuous-spin field in $AdS_{d+1}$ is labeled by the parameters $e_\imrm$, $s_\imrm$, $\sbf^\bfrm$. The $e_\imrm$ and $s_\imrm$ are purely imaginary numbers. The $\sbf^\bfrm$ defined in \rf{26042021-manus-01-a5}-\rf{26042021-manus-01-a6x} labels irrep of the $so(d-2)$ algebra. Note that the labels $e_\imrm$ and $s_\imrm$ are related to the commonly used labels $E_0$ and $s$ as
\be
E_0 = \frac{d}{2} + e_\imrm\,, \qquad s = \frac{2-d}{2} + s_\imrm\,.
\ee

\noinbf{ii}) The spin operators \rf{27042021-manus-23},\rf{27042021-manus-24} are polynomials in the vector variable $\zeta^i$ and degree-2 polynomials in the derivative $\partial/\partial \zeta^i$.

\noinbf{iii}) Using constraints \rf{26042021-manus-01-a3}-\rf{26042021-manus-01-a12}, we get the description of the irreducible mixed-symmetry continuous-spin fields, while, ignoring  those constraints, we get the description of the triplectic mixed-symmetry continuous-spin fields. Thus our formulation, among other things, provides the easy and quick access to the triplectic mixed-symmetry continuous-spin AdS fields.

\medskip

\noindent {\bf Conclusions}. In this paper, we developed the light-cone formulation of the bosonic and fermionic mixed-symmetry continuous-spin fields in AdS and flat spaces. In our formulation, the spin operators are polynomial in the vector variable $\zeta^i$ and the derivative $\partial/\partial \zeta^i$. This will provide considerable simplifications in the study of interacting vertices for the continuous-spin fields. In Refs.\cite{Metsaev:2005ar,Metsaev:2007rn},
we used light-cone approach for the study of interacting bosonic and fermionic mixed-symmetry  discrete-spin fields in the flat space, while, in Refs.\cite{Metsaev:2017cuz,Metsaev:2018moa}, we used  light-cone approach for the study of interacting bosonic continuous-spin fields in the flat space. We believe therefore that the methods in Refs.\cite{Metsaev:2017cuz,Metsaev:2018moa,Metsaev:2005ar,Metsaev:2007rn} with the result in this paper will provide the possibilities for the study of interacting mixed-symmetry continuous-spin fields in the flat space.
Application of our result for the study of interacting continuous-spin AdS fields could be of great interest. In this respect we think  that method for the study of interacting discrete-spin light-cone gauge AdS fields developed in \cite{Metsaev:2018xip} and the light-cone bootstrap in Ref.\cite{Skvortsov:2018uru} will be helpful for the study of interacting continuous-spin AdS fields. Lorentz covariant methods for the study of mixed-symmetry discrete-spin AdS fields may be found in Refs.\cite{Alkalaev:2010af,Boulanger:2011qt,Boulanger:2011se}. Application of BRST method to the problem of interacting discrete-spin fields may be found in Refs.\cite{Fotopoulos:2008ka,Metsaev:2012uy,Grigoriev:2020lzu}.
Other interesting methods for the study of interacting discrete-spin fields in AdS and flat spaces may be found in Refs.\cite{Vasilev:2011xf,Manvelyan:2010jr}. Studies of  supersymmetric interacting discrete-spin fields in AdS and flat spaces by using various methods may be found in Refs.\cite{Alkalaev:2002rq,Metsaev:2019dqt,Metsaev:2019aig}.
Investigation of the mixed-symmetry continuous-spin fields along the line of the group-theoretical methods in Refs.\cite{Basile:2017kaz,Khan:2021dgj} could also be of some interest.

\bigskip
{\bf Acknowledgments}. This work was supported by the RFBR Grant No.20-02-00193.

\vspace{-0.3cm}
%%%%%%%%%%%%%%%%%%%%%%%%%%%%%%%%%%%%%%%%%%%%%%%%%%%%%%%%%%%%%%%%%%%%%%
\setcounter{section}{0} \setcounter{subsection}{0}
%%%%%%%%%%%%%%%%%%%%%%%%%%%%%%%%%%%%%%%%%%%%%%%%%%%%%%%%%%%%%%%%%%%%%%

\end{document}